# Metasurface Holography over 90% Efficiency in the Visible via Nanoparticle-Embedded-Resin Printing

Joohoon Kim,[1,†] Dong Kyo Oh,[1,†] Hongyoon Kim,[1] Gwanho Yoon,[2] Chunghwan Jung,[3] Jae Kyung Kim,[1] Trevon Badloe,[1] Seokwoo Kim,[1] Younghwan Yang,[1] Jihae Lee,[3] Byoungsu Ko,[1] Jong G. Ok,[4] and Junsuk Rho[1,3,5,6,*]

[1]Department of Mechanical Engineering, Pohang University of Science and Technology (POSTECH), Pohang 37673, Republic of Korea

[2]Department of Manufacturing Systems and Design Engineering, Seoul National University of Science and Technology, Seoul 01811, Republic of Korea

[3]Department of Chemical Engineering, Pohang University of Science and Technology (POSTECH), Pohang 37673, Republic of Korea

[4]Department of Mechanical and Automotive Engineering, Seoul National University of Science and Technology, Seoul 01811, Republic of Korea

[5]POSCO-POSTECH-RIST Convergence Research Center for Flat Optics and Metaphotonics, Pohang 37673, Republic of Korea

[6]National Institute of Nanomaterials Technology (NINT), Pohang 37673, Republic of Korea

[†]These authors contributed equally to this work.

*Corresponding author. E-mail: jsrho@postech.ac.kr

## Abstract

Metasurface holography, the reconstruction of holographic images by modulating the spatial amplitude and phase of light using metasurfaces, has emerged as a next-generation display technology. However, conventional fabrication techniques used to realize metaholograms are limited by their small patterning areas, high manufacturing costs, and low throughput, which hinder their practical use. Herein, we demonstrate a high efficiency hologram using a one-step nanomanufacturing method with a titanium dioxide nanoparticle-embedded-resin, allowing for high-throughput and low-cost fabrication. At a single wavelength, a record high 96.4% theoretical efficiency is demonstrated with an experimentally measured conversion efficiency of 90.6% and zero-order diffraction of 7.3% producing an ultrahigh-efficiency, twin-image free hologram, that can even be directly observed under ambient light conditions. Moreover, we design a broadband meta-atom with an average efficiency of 76.0% and experimentally demonstrate a metahologram with an average efficiency of 62.4% at visible wavelengths from 450 to 650 nm.

## 1. Introduction

Metasurfaces[1-3] composed of arrays of subwavelength structures have been intensively researched due to their ability to modulate electromagnetic (EM) waves. Various practical applications of metasurfaces have been reported, such as metalenses,[4-6] color filters,[7-11] beam modulating devices,[12,13] and all solid-state light detection and ranging (LiDAR).[14,15] Moreover, high-quality, high-resolution, and large field of view (FOV) holograms have been demonstrated.[16-23]

Metaholograms are designed to reconstruct images by manipulating the spatial amplitude and phase of light using metasurfaces, and have emerged as a next-generation display technology due to their compact form factor. In particular, many studies have been conducted to increase the efficiency of metaholograms for practical applications.[24] Although plasmonic metaholograms generally exhibit low efficiency due to Ohmic losses, a reflection-type plasmonic metahologram reaching up to 80% efficiency at 825 nm has been successfully demonstrated.[17] All-dielectric metasurfaces are another candidate to achieve high efficiency metaholograms. An all-dielectric Huygens' metahologram consisting of silicon (Si) meta-atoms has been demonstrated with a transmission efficiency of 86% at 785 nm.[25] Additionally, amorphous titanium dioxide (a-$TiO_2$) has been used to produce metaholograms reaching efficiencies of 82%, 81%, and 78% at wavelengths of 480, 532, and 660 nm, respectively.[18]

Conventional fabrication techniques for realizing all-dielectric metaholograms such as electron-beam lithography (EBL) and focused ion beam (FIB) milling, to name a few, have inherent limitations due to their small pattern areas, high cost, and low throughput, which are the biggest restrictions on commercializing metaholograms.[26] Extreme ultraviolet (EUV) lithography can be used to fabricate nanostructures on a large scale,[27-29] but is, unfortunately, still extremely expensive. To overcome these limitations, nanoimprint lithography (NIL) has been introduced as a way to easily and quickly replicate nanostructures from a master mold directly onto a substrate.[30-32] Since the imprinted resin generally acts as an intermediate structure, conventional NIL also requires additional operations such as deposition and etching in order to complete the fabrication of the final metasurface. Although the refractive index of typical resin is typically too low for use in metasurfaces,[33] it has successfully been used to build meta-atoms[34] by utilizing extremely high-aspect-ratio structures. Recently, the single-step manufacturing of a metalens using NIL and a new type of nanoparticle-embedded-resist (nano-PER) has been demonstrated. The refractive index of the nano-PER can be increased at will by embedding nanoparticles (NPs) of different materials into a standard resin, which opens up exciting new possibilities of using the resin directly as functional meta-atoms.[35-37] However, conventional metasurfaces consisting of nano-PER showed a low efficiency 33% in the visible regime, resulting in a reduction of performance and applicability.[36]

Here, we introduce a metahologram made of a nano-PER reaching a record high of 96.4% efficiency, fabricated using a high-throughput one-step NIL manufacturing process. We verify that the optical properties of the nano-PER are sufficient to produce high efficiency metaholograms and even outperform other low-loss dielectric materials. For the synthesis of the nano-PER, titanium dioxide ($TiO_2$) NPs, that have a high refractive index in the visible regime, are dispersed in an ultraviolet (UV)-curable resin. Using a single step of NIL, the meta-atom patterns are transferred from a master mold and cured through UV light to directly imprint the metasurface without the need of any secondary operations. We theoretically verify that the metahologram composed of the $TiO_2$ nano-PER can achieve a 96.4% conversion efficiency while being almost free of zero-order diffraction (0.2%) at a wavelength of $\lambda = 532$ nm. We confirm this high performance by experimentally demonstrating a metahologram with a measured conversion efficiency of 90.6% with no twin image. Furthermore, we reveal the broadband property of $TiO_2$ nano-PER by designing a single meta-atom with high efficiency over the visible regime. We theoretically calculate an average efficiency of 76. 0% and experimentally demonstrate a metahologram with an average efficiency of 62.4% for different wavelengths from 450 to 650 nm. This low-cost and high-throughput realization of high-efficiency metaholograms could prove to be a valuable method for the commercialization of optical metasurfaces.

## 2. Methods and results

*2.1 Optical properties of $TiO_2$ nano-PER*

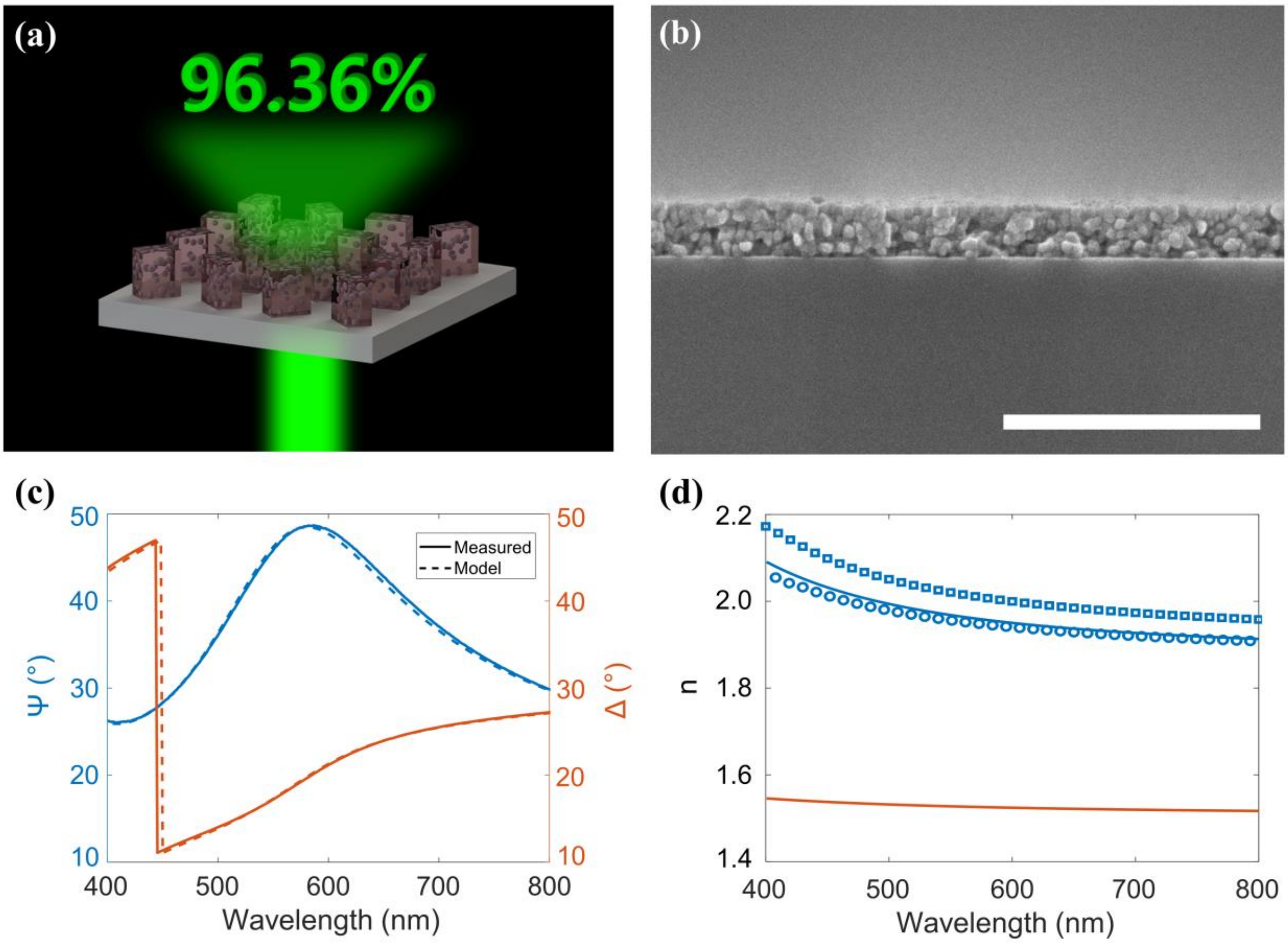

**Figure 1.** Optical properties of $TiO_2$ nanoparticle-embedded-resin (nano-PER). (a) Schematic of the high-efficiency metahologram made up of nano-PER meta-atoms. (b) Scanning electron microscope (SEM) image of a cross-section of the spin-coated nano-PER films with a $TiO_2$ weight ratio of 80%. Scale bar: 1 μm (c) Ellipsometry analysis of the $TiO_2$ nano-PER film. (d) The measured refractive index of a typical resin (orange line) and the nano-PER film with 80% weight ratio (blue line); blue circles: calculated data by the Maxwell-Garnett formula; blue squares: simulated data by Comsol Multiphysics version 5.6.

The key to using NIL as a one-step fabrication method is to utilize the printable resin as the desired meta-atom directly (**Figure 1a**). Typically, resins used in NIL have a refractive index (*n*) of around ~1.5, and therefore cannot be used as meta-atoms in optical metasurfaces without extremely large structure parameters. In order to increase the *n* of the resin in the visible regime, we use the well-understood and inexpensive synthesis of $TiO_2$ NPs[38] to create a $TiO_2$ nano-PER. This is done by dispersing 30 nm diameter $TiO_2$ NPs into a UV-curable resin, which acts as a matrix to enable the direct transfer of nanostructures from the master mold. The result is a homogeneous effective medium with an increased effective index, closer to that of $TiO_2$ than the bare resin.

The optical properties of the nano-PER are determined by the weight ratio of the NPs. Here, we use a $TiO_2$ nano-PER with a weight ratio of 80%. To determine the $n$ and extinction coefficient ($k$) of the $TiO_2$ nano-PER, an 84.9 nm thin film of $TiO_2$ nano-PER was spin-coated onto a Si substrate (**Figure 1b**) and the amplitude ratio ($\Psi$) and phase difference ($\Delta$) between the $s$ and $p$ components of the thin film were measured using ellipsometry (**Figure 1c**). The measured $\Psi$ and $\Delta$ data are fitted using the Cauchy dispersion model, which is expressed as follows[39]

$$n(\lambda) = A + \frac{B}{\lambda^2} + \frac{C}{\lambda^4} + \cdots, \tag{1}$$

where $n(\lambda)$ is the dispersive refractive index, $\lambda$ is the wavelength of the light, and $A$, $B$, $C$ are coefficients to fit the model with the measured $\Psi$ and $\Delta$ data. Model coefficients of $A = 1.87600$, $B = 0.02025$, and $C = 0.00216$ were determined, with a mean squared error (MSE) of 19.070. This confirms the validity of treating the $TiO_2$ nano-PER as a homogeneous effective medium. The measured $n$ of 80% weight ratio $TiO_2$ nano-PER and typical resin at a wavelength of 532 nm is 1.9754, and 1.5288, respectively (**Figure 1d**). $TiO_2$ nano-PER with a weight ratio of 80% has a higher refractive index than typical resin while $k$ remains near zero (**Supplementary Data 1**). Furthermore, we measure and compare the refractive index with different $TiO_2$ weight ratios and confirm that the refractive index increases as the weight ratio increases (**Supplementary Data 2**).

We also calculated the effective refractive index (n) of the nano-PER with a $TiO_2$ filling ratio of 80% using the following Maxwell-Garnett mixing formula.[40]

$$n_{MG} = \sqrt{\varepsilon_r \frac{1+2f\alpha}{1-f\alpha}}, \tag{2}$$

where $\alpha = (\varepsilon_{TiO_2} - \varepsilon_r)/(\varepsilon_{TiO_2} + 2\varepsilon_r)$; f is the volume fraction of $TiO_2$ and $\varepsilon_{TiO_2}$ , $\varepsilon_r$ are the permittivity of $TiO_2$ and resin, respectively. Using the Maxwell-Garnett mixing formula, we estimate the volume fraction to be 44% for $TiO_2$ weight ratio of 80% (**Figure 1d**). The simulated refractive index from COMSOL Multiphysics version 5.6 using the calculated filling ratio agrees well with the measured refractive index indicating the credibility of the estimated filling ratio. Furthermore, this is also confirmed by the Bruggeman mixing formula (**Supplementary Data 3**).

Compared to typical resins, the increased *n* of the $TiO_2$ nano-PER provides the ability to confine light much more strongly. Additionally, the extremely low *k* indicates that the $TiO_2$ nano-PER has negligible absorption, and therefore almost zero optical loss at visible wavelengths. These properties reinforce the choice of material for use in metasurfaces that operate in the visible regime, particularly for highly efficient metaholograms.

### *2.2 Design of high-efficiency meta-atoms*

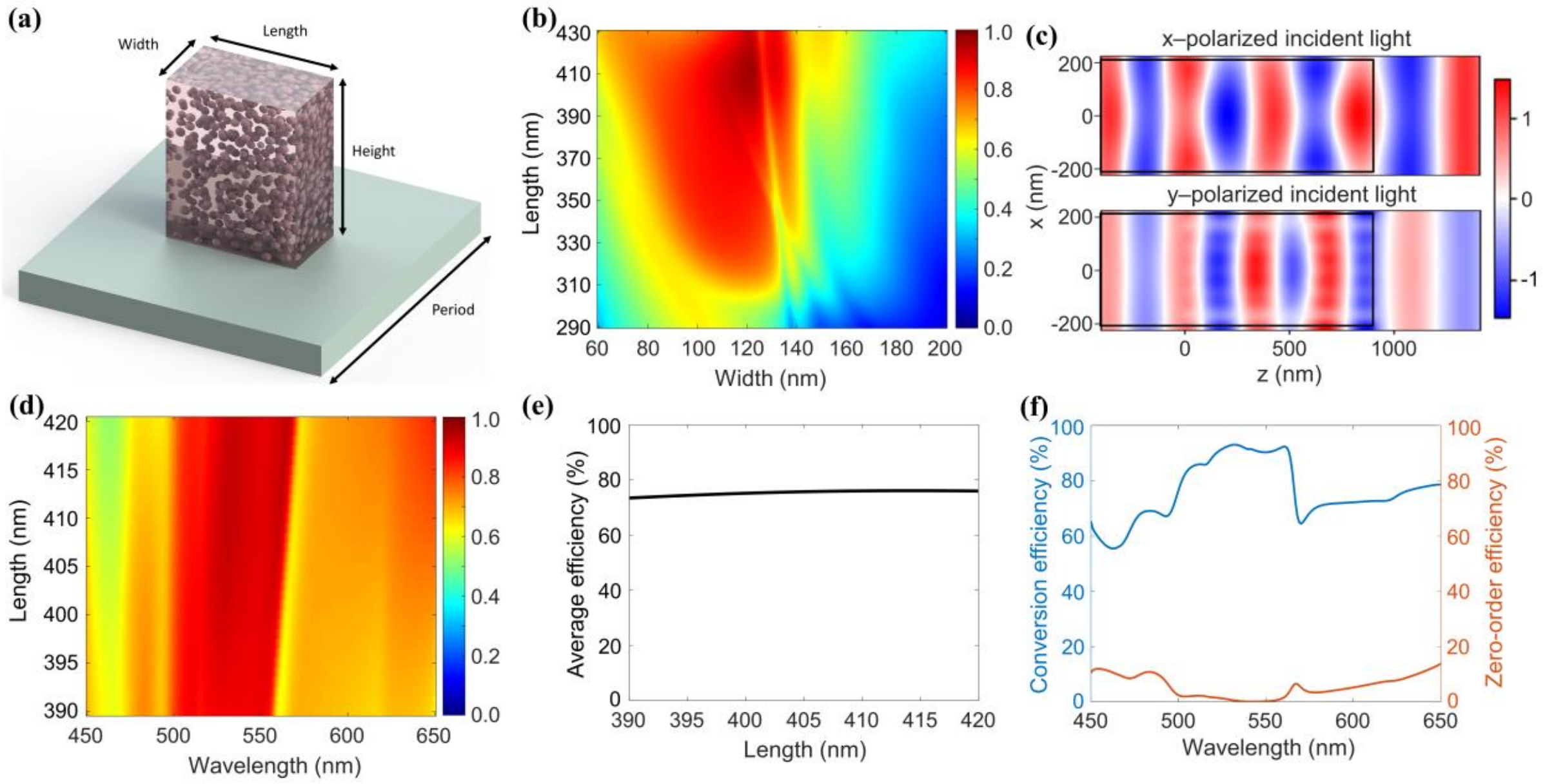

**Figure 2.** Design and simulation of the high-efficiency meta-atom. (a) Configuration of the $TiO_2$ nano-PER meta-atom structure. (b) The calculated conversion efficiency of the meta-atom at $\lambda = 532$ nm with fixed height (910 nm) and period (450nm). (c) Simulated electric field profiles for *x*- (above) and *y*-polarized incident light (below). (d) The calculated conversion efficiency for the broadband property with fixed width (116 nm), height (940 nm), and period (450 nm). (e) Average efficiency for wavelength from 450 to 650 nm. (f) Simulated optical efficiency for both the conversion and zero-order diffraction.

Rigorous coupled-wave analysis (RCWA)[41] was used to simulate the optical properties of the $TiO_2$ nano-PER-based meta-atoms using the measured *n*. Anisotropic meta-atoms were designed to modulate the incident light using the concept of Pancharatnam-Berry phase (PB phase), also known as geometric phase, (**Figure 2a**) to allow for broadband operation. PB phase achieves a full $2\pi$ phase modulation by modifying the in-plane orientation angle of the anisotropic meta-atoms in the metasurface and can be understood by analyzing the meta-atoms with a Jones matrix **J**,[42] given by

$$\mathbf{J} = \begin{bmatrix} t_{xx} & 0 \\ 0 & t_{yy} \end{bmatrix}, \quad (3)$$

where $t_{xx}$ and $t_{yy}$ represent the complex transmission coefficients for light polarized along the long and short axes of the anisotropic meta-atoms, respectively. The Jones matrix **T** for meta-atoms rotated by an angle $\theta$ can be calculated using a rotation matrix $\mathbf{R}(\theta)$, as follows[19]

$$\mathbf{T} = \mathbf{R}(-\theta)\ \mathbf{J}\ \mathbf{R}(\theta). \quad (4)$$

The transmission coefficients for rotated meta-atoms can then be calculated, as follows

$$E_T = \mathbf{T}E_{CP} = \frac{t_{xx}+t_{yy}}{2}E_{co-pol} + \frac{t_{xx}-t_{yy}}{2}e^{i2\theta}E_{cross-pol}, \quad (5)$$

where $E_T$, $E_{CP}$, $E_{co\text{-}pol}$, and $E_{cross\text{-}pol}$ represent the electric fields of the transmitted light, the circularly polarized incidence, and the co- and cross-polarized components, respectively. The $e^{i2\theta}$ term indicates that the phase of the cross-polarized part of the transmitted light is retarded by $2\theta$, and depends solely on the orientation angle of the meta-atom. The co-polarized component of the light causes zero-order diffraction, which leads to a low signal-to-noise ratio and poor-quality holograms. Therefore, the conversion efficiency, which is defined as the amplitude of the cross-polarized component, is directly connected to the efficiency of the hologram. To demonstrate high-quality holograms, the conversion efficiency should be maximized, while the co-polarized light is simultaneously suppressed.

We calculate the conversion efficiency of the meta-atoms by sweeping lengths ($l$) from 290 to 430 nm and widths ($w$) from 60 to 200 nm, with a fixed height ($h$) of 910 nm and periodicity ($p$) of 450 nm (**Figure 2b**). $h$ is determined to produce the maximum efficiency (**Supplementary Data 4**), while $p$ is chosen to be smaller than the operating wavelength (532 nm) to suppress the diffraction of light. The meta-atom with $l$ = 411 nm and $w$ = 121 nm reaches up to the 96.4% conversion efficiency with almost zero (0.2%) zero-order diffraction at a wavelength of 532 nm, while only absorbing 0.2% due to the extremely low $k$ of the $TiO_2$ nano-PER. Considering that reflection inevitably occurs when light passes through the interface of media with different refractive indices, the reflection should be reduced to achieve high efficiency. We design the metasurface with index matching in mind so it acts as an anti-reflective (AR) film, along with negligible absorption from the $TiO_2$ nano-PER, thus achieving high efficiency with only 3.3% reflection (**Supplementary Data 5**). Moreover, many candidate structures have an efficiency of over 90% near the target structure, therefore some fabrication errors are acceptable. To confirm that the $l$ and $w$ of the chosen meta-atom provides a $\pi$–phase difference between the $x$ and $y$ components of the electric field ($\mathbf{E_x}$ and $\mathbf{E_y}$), we simulate the propagating electric field profiles of $x$- and $y$-polarized light in the optimized meta-atom at 532 nm (**Figure 2c**). The results clearly show that the meta-atom acts as a half-wave plate, therefore producing the maximum efficiency.

Furthermore, we reveal the broadband properties of the nano-PER based metahologram. To find the meta-atom that maximizes the broadband properties, we calculate the conversion efficiency of meta-atoms with lengths from 390 to 420 nm, and a fixed width of 116 nm, height of 940 nm, and periodicity of 450 nm (**Figure 2d**). The meta-atoms show high efficiency over the broadband visible range from 450 nm to 650 nm. Moreover, we calculate the average efficiency from 450 nm to 650 nm along each length (**Figure 2e**). As the length increases, the average efficiency increases from 73.4% (390 nm) to 75.9% (420 nm), and fabrication errors are acceptable because meta-atoms show broadband property with high efficiency over a wide range of lengths. The conversion efficiency and zero-order diffraction efficiency of the meta-atom with a length of 410 nm is calculated over the entire visible regime (**Figure 2f**) to confirm o the high efficiency broadband properties over the visible regime.

### *2.3 One-step nanofabrication using nano-PER for high-efficiency metaholograms*

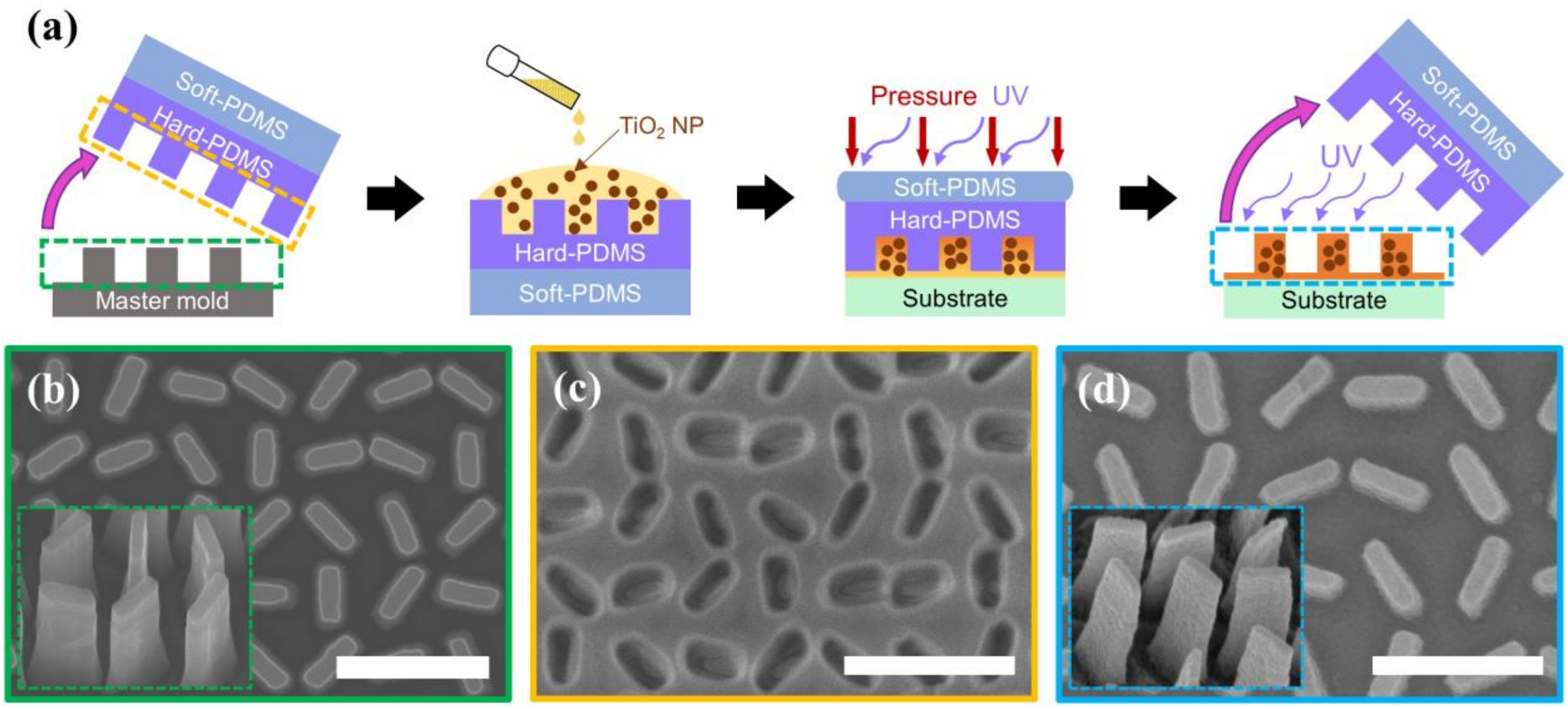

**Figure 3.** Fabrication of the $TiO_2$ nano-PER metahologram. (a) Schematic of the one-step fabrication process. (b) SEM image of the master mold fabricated by EBL on the Si substrate. (Inset) tilted view. (c) SEM image of the soft mold composed of *h*-PDMS and PDMS composite with the metasurface. (d) SEM image of the final structure of the printed $TiO_2$ nano-PER. (Inset) titled view. All scale bars: 1 μm

A schematic diagram of the fabrication process is provided in **Figure 3a**. First, a master mold with the desired metasurface is fabricated on a Si substrate using traditional EBL, lift-off, and etching processes (details in the Experimental Section). **Figure 3b** shows a scanning electron microscope (SEM) image of the Si master mold. The master mold is hydrophobically coated with a vaporized fluorine solution for the subsequent demolding process of a soft mold. Next, A hard-polydimethylsiloxane ($h$-PDMS) solution is coated and solidified on the master mold to replicate the metasurface precisely. The $h$-PDMS is preferred as it can transfer nanopatterns at a higher resolution than standard polydimethylsiloxane (PDMS) due to its high modulus (～9 $N/mm^2$).[43] The PDMS is then coated onto the $h$-PDMS layer to adjust the thickness of the soft mold, which is important because if the soft mold is too thin, it will shrink overly while absorbing the solvent in the nano-PER. The PDMS layer is then heated and combined with the h-PDMS layer, and the soft mold is manufactured by separating the h-PDMS layer from the master mold (**Figure 3c**). After spin-coating the nano-PER with $TiO_2$ NPs on the soft mold, the coated nano-PER contacts uniformly with the glass substrate. Finally, a correct amount of pressure and UV light are applied to the glass-PER-soft mold structure. When the nano-PER is fully cured, the soft mold is detached, leaving the high-aspect-ratio metasurface on the substrate, as can be seen in **Figure 3d**.

This single-step process of NIL with nano-PER is very attractive as it not only allows for the high-throughput fabrication of metasurfaces but also can be easily complemented by enhancing or weakening the work of adhesion ($W_a$) between interfaces to fabricate high-efficiency metaholograms. While the soft mold is hydrophobically coated to decrease the $W_a$ between the soft mold and the nano-PER, the PMMA layer is coated on the pre-treated glass substrate by $O_2$ plasma simultaneously to increase the interaction between the nano-PER and the glass substrate, which facilitating the replication of metasurfaces with an aspect ratio of about 8. The $W_a$ between two interfaces can be calculated by harmonic-mean equation,[44,45] given by:

$$W_a = 4\left(\frac{\gamma_{s1}^d \gamma_{s2}^d}{\gamma_{s1}^d + \gamma_{s2}^d} + \frac{\gamma_{s1}^p \gamma_{s2}^p}{\gamma_{s1}^p + \gamma_{s2}^p}\right), \tag{6}$$

where the subscript s is for the interface of solid and the superscripts d and p are for the dispersion and polar components of the surface tension $\gamma$, respectively. At this point, surface tension is determined by the geometric-mean equation,[44,45] given by:

$$(1 + \cos\theta_l)\gamma_l = 2(\gamma_s^d \gamma_l^d)^{\frac{1}{2}} + 2(\gamma_s^p \gamma_l^p)^{\frac{1}{2}}, \tag{7}$$

where $\theta$ is for the contact angle of the liquid on the solid and subscript l and s are for the liquid and solid components of the $\gamma$, respectively. By solving simultaneous equations about two different liquids which know the dispersion and polar surface tensions, typically deionized (DI) water and diiodomethane, we can decide dispersion and polar surface tensions of various surfaces and calculate the $W_a$ (**Supplementary Data 6**). While the bare glass substrate without pre-treatment has the $W_a$ about 62.7 mJ/m$^2$, the PMMA layer-coated glass has the $W_a$ about 65.9 mJ/m$^2$. To increase the surface energy of interfaces more, we adopt additional $O_2$ plasma treatment, so the PMMA layer on the $O_2$ plasma-treated glass has $W_a$ about 66.5 mJ/m$^2$ with $TiO_2$ PER, whereas only $O_2$ plasma-treated glass has a lower $W_a$ rather. Likewise, the $W_a$ of the fluor-coated soft mold becomes lower than the non-treated soft mold, presenting 39.2 and 47.6 mJ/m$^2$, respectively. Consequently, the pre-treated substrate and soft mold have $W_a$ difference about 27.3 mJ/m$^2$, but the non-treated interfaces have only $W_a$ difference about 15.1 mJ/m$^2$. By increasing $W_a$ difference between the substrate and soft mold about two times, we can make metasurfaces with a high aspect ratio over 8, straightforwardly.

*2.4 Design and demonstration of a high-efficiency hologram*

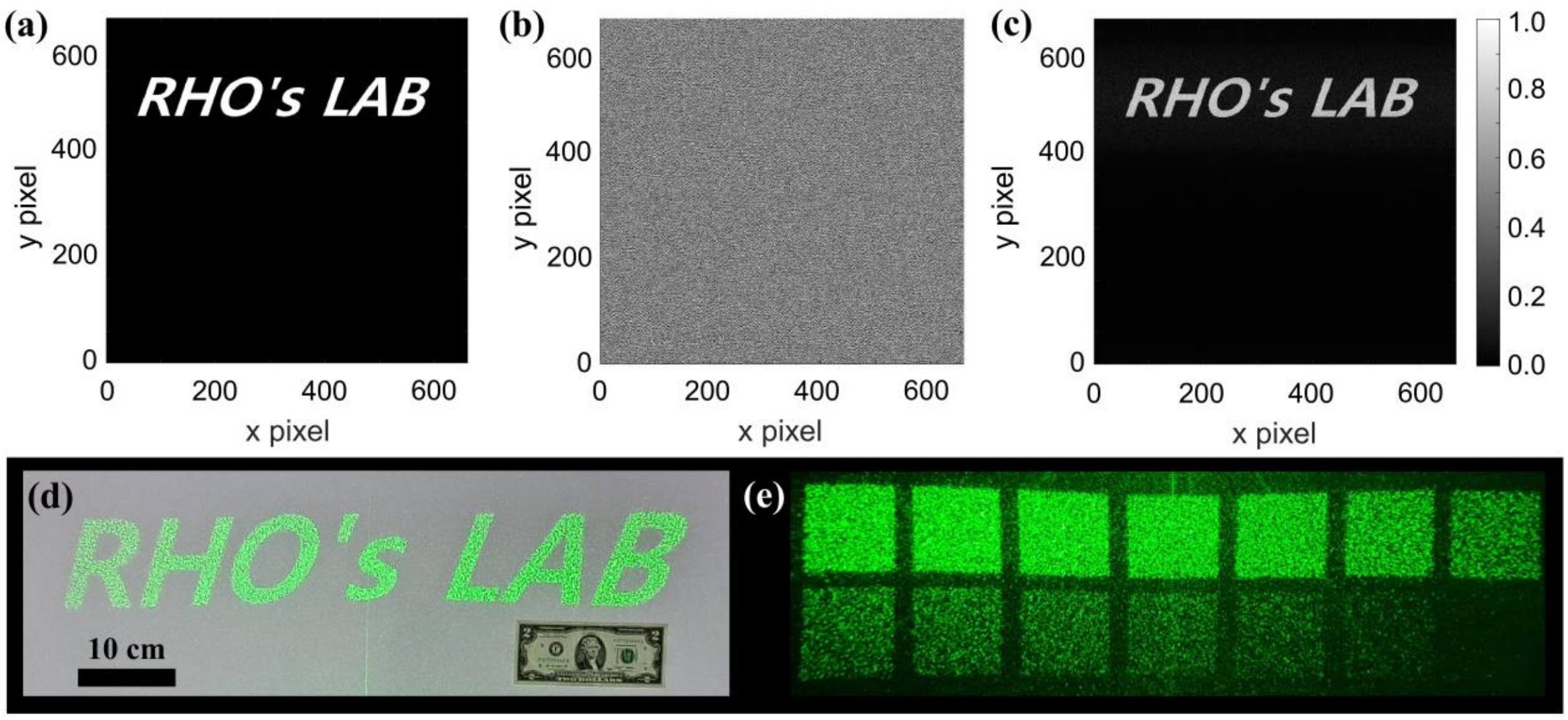

**Figure 4.** Design and experiment of the high-efficiency metahologram. (a) Designed holographic image. (b) Phase map of the computer-generated Fraunhofer hologram. (c) Simulated hologram. (d) Captured image of the hologram when illuminated with a 532 nm wavelength laser under ambient lighting conditions. A 2-dollar bill is pictured next to the holographic image for scale. (e) Captured image of the gradient palette hologram with conversion efficiency of 96.4%.

We designed a simple Fraunhofer hologram with the name of our research group, "RHO's LAB", as the holographic image to be reconstructed in the far-field (**Figure 4a**). The Gerchberg-Saxton (GS) algorithm was used to retrieve a high-quality phase-only hologram.[46] The recovered hologram exhibits a pincushion-like distortion due to the property of the Fraunhofer approximation. We compensate for the distortion using barrel distortion and retrieve a phase map of the hologram using the Gerchberg-Saxton (GS) algorithm (**Figure 4b**). Since the phase is only dependent on the angle of rotation of the meta-atoms, we discretize the phase into 16 steps in order to limit the noise added to the hologram at this step. We confirm that the phase map is calculated well and 16 steps of discretized phase are enough to recover the hologram by recreating the holographic image from the discretized phase-only hologram (**Figure 4c)**. By modulating the phase with the designed meta-atom, we experimentally realize a high-efficiency hologram using $TiO_2$ nano-PER (**Figure 4d**). In order for metaholograms to be used in general lighting conditions, it is important that the holographic images appear clearly even under ambient lighting. The low efficiency of conventional metaholograms is a limiting factor for achieving this. Furthermore, it is hard to display large holographic images with low efficiency holograms since the intensity of light per unit area decreases as the hologram size increases. The far-field holographic image of our metahologram is clear, vivid, free from twin images, and over 60 cm in size. The high-efficiency means that it is also visible under ambient lighting, proving its validity for practical applications.

High efficiency holograms have the advantage of being able to demonstrate various steps of brightness. Low efficiency holograms are unable to achieve this due to the small difference in contrast between the bright and dark shades. To experimentally verify this, we confirm that various steps of brightness are possible by implementing a gradient palette with high efficiency holograms with different brightness (**Figure 4e**) and demonstrate that our high efficiency holograms have a much better performance compared to gradient palette holograms with the efficiencies of 70%, 50% and 30% (**Supplementary Data 7**). A schematic of the experimental setup used to reconstruct the hologram is provided in **Supplementary Data 8**. A linear polarizer (LP) and a quarter-wave plate (QWP) are used to produce circularly polarized light, which is incident on the sample and is converted to the opposite handedness by the metasurface, generating a holographic image. A standard plastic viewing screen was used to observe the hologram in the far-field.

To determine the efficiency of the $TiO_2$ nano-PER hologram, we experimentally measured the conversion efficiency. A schematic of the experimental setup and a detailed explanation of the calculating efficiency is provided in **Supplementary Data 9**. The light diffracts during the propagation through the sample and is focused through the lens. The focused light passes through the quarter-wave plate and linear polarizer. Depending on the rotation of the quarter-wave plate, only the intensity of the converted or zero-order light is measured by the photodetector. The measured conversion efficiency with 532 nm laser is a record 90.6%, with a measured zero-order efficiency of 7.3%, which closely matches with the theoretically calculated values.

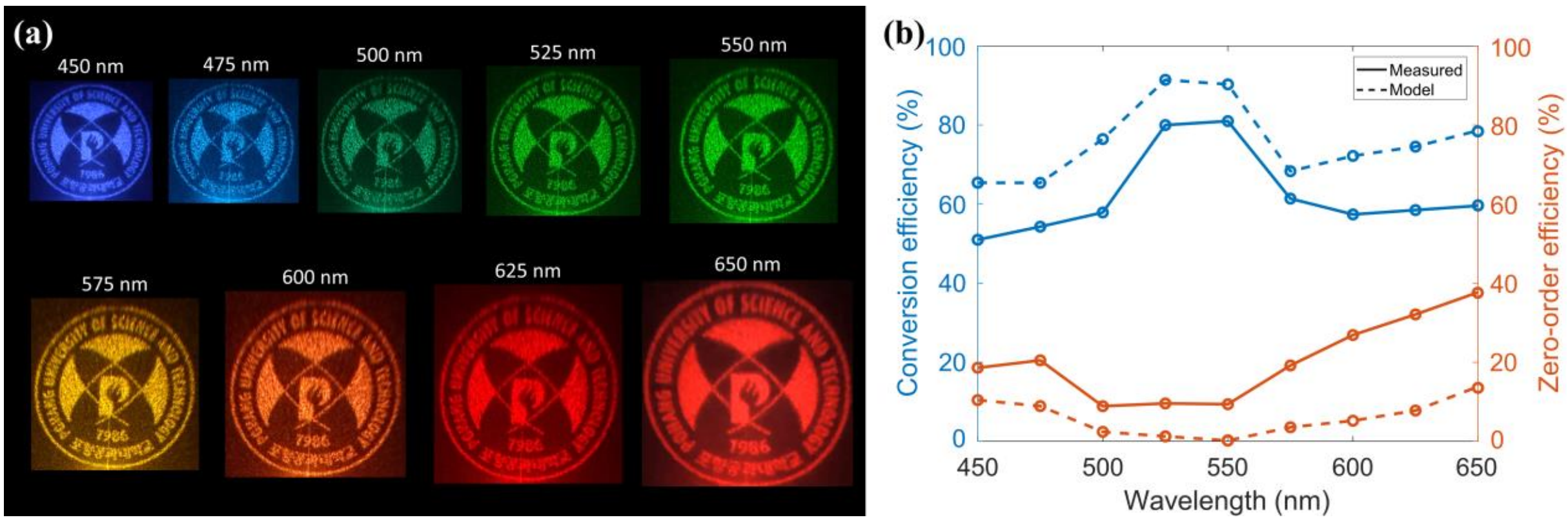

**Figure 5.** Broadband property of the high-efficiency metahologram covering the visible spectrum. (a) Holographic images with different visible wavelength inputs. All images are obtained from proposed meta-atom with broadband property. (b) Measured (solid line) and simulated (dashed line) data of conversion (blue) and zero-order (orange) efficiency.

Moreover, we experimentally verify the broadband properties of $TiO_2$ nano-PER based metahologram by reconstructing high efficiency holograms with lasers of various wavelengths from 450 to 650 nm (**Figure 5a**). We measure and calculate the conversion efficiencies of each demonstrated holographic image (**Figure 5b**). We experimentally confirm that the designed broadband metahologram has an average efficiency of 62.4% for nine different wavelengths from 450 to 650 nm. The measured and theoretical efficiencies match and prove the excellent broadband property. Finally, we confirm that our work has both higher theoretical and experimental efficiencies compared to previously reported $TiO_2$ nano-PER metasurfaces (**Figure 6**) and high efficiency metaholograms and metalenses, particularly in the visible regime (**Supplementary Data 10**).

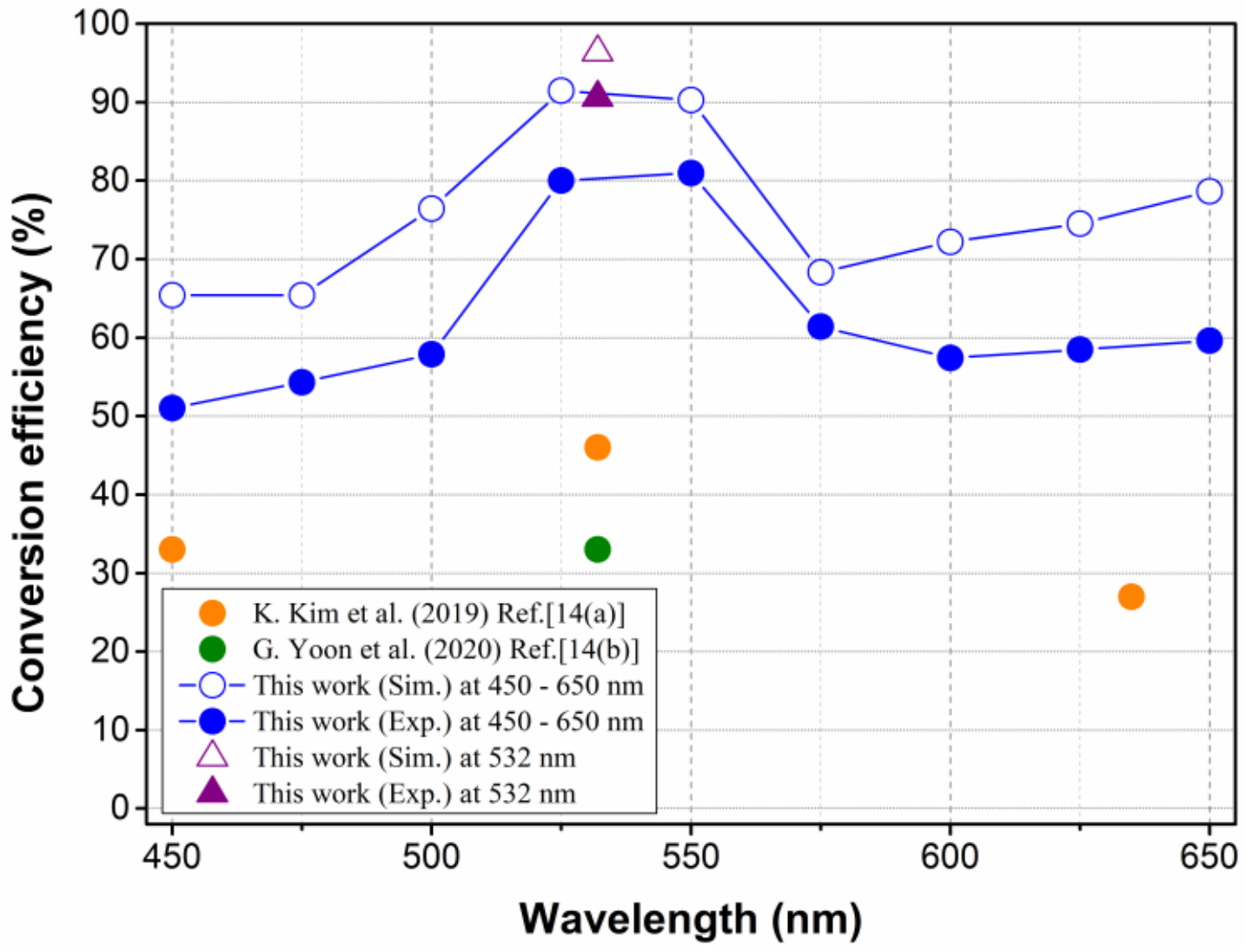

**Figure 6.** Direct comparison of conversion efficiency of $TiO_2$ nano-PER based metasurfaces.

## 3. Conclusion

In summary, we have demonstrated an almost perfectly noise-free hologram, reaching, to the best of our knowledge, a record high of 96.4% theoretical efficiency and 90.6% experimental efficiency at a wavelength of 532 nm using a one-step manufacturing method. We synthesized the $TiO_2$ nano-PER by combining $TiO_2$ NPs with a typical nanoimprint resin to increase its refractive index to allow it to be used directly as a meta-atom for use in metasurfaces that operate in the visible regime. We numerically designed and characterized a metahologram with a conversion efficiency of 96.4% and zero-order efficiency of 0.2% at a wavelength of 532 nm. We experimentally demonstrated the metahologram and measured values of 90.6% conversion efficiency and 7.3% zero-order efficiency, respectively. Due to the extremely high efficiency of the hologram, the designed holographic image was clearly visible even under ambient lighting conditions, proving the applicability of our device in holographic displays and augmented reality devices. Moreover, we reveal the broadband property of $TiO_2$ nano-PER by designing a meta-atom which has an average efficiency of 76.0% from 450 to 650 nm. We confirm that experimentally demonstrated metahologram has an average efficiency of 62.4% for various wavelengths from 450 to 650 nm.

Our metahologram made up of nanoimprinted nano-PER structures was fabricated using a reusable soft mold. This method of producing high-efficiency metaholograms has the added benefits of high-throughput and low fabrication cost. Specifically, each sample takes around 15 minutes and costs less than 1.4 USD to realize (**Supplementary Data 11**). Given these obvious benefits of using nano-PER with NIL, we believe this work could lead to the commercialization of metaholograms. Finally, we note that multifunctional metaholograms could be fabricated by integrating the nano-PER meta-atoms with conventional tuning techniques,[47] thus proving their promise as candidates for holographic display technology, as well as in augmented and virtual reality devices.

## 4. Materials and methods

*Synthesis of $TiO_2$ nano-PER*: The $TiO_2$ nano-PER was prepared by mixing $TiO_2$ NPs dispersed in MIBK (DT-TIOA-30MIBK (N30), Ditto technology), monomer (dipentaerythritol penta-/hexa- acrylate, Sigma-Aldrich), photo-initiator (1-Hydroxycyclohexyl phenyl ketone, Sigma-Aldrich), and MIBK solvent (MIBK, Duksan general science). The mixing ratio was controlled to achieve a weight ratio of 4 wt % for $TiO_2$ NPs, 0.7 wt % for monomer, and 0.3 wt % for photo-initiator.

*Fabrication of the master mold*: A Si substrate was used for the master mold. The meta-atoms were transferred onto a bilayer of two positive tone photoresists (495 PMMA A6, MicroChem & 950 PMMA A2, MicroChem) using the standard EBL process (ELIONIX, ELS-7800; acceleration voltage: 80 kV, beam current: 100 pA). The exposed patterns were developed by MIBK/IPA 1:3 developer mixed solution. An 80 nm-thick chromium (Cr) layer was deposited using electron beam evaporation (KVT, KVE-ENS4004). The lifted-off Cr meta-atoms were used as an etching mask for the Si substrate. Cr patterns were transferred onto the Si substrate using a dry etching process (DMS, silicon/metal hybrid etcher). The remaining Cr etching mask was removed by Cr etchant (CR-7).

*Fabrication of the soft mold*: *h*-PDMS was prepared by mixing 3.4 g of vinylmethyl copolymers (VDT-731, Gelest), 18 μL of platinum-caralyst (SIP6831.2, Gelest), 0.1 g of the modulator (2,4,6,8- tetramethyl-2,4,6,8-tetravinylcyclotetrasiloxane, Sigma-aldrich), 2 g of toluene, and 1 g of siloxane-based silane reducing agent (HMS-301, Gelest). The *h*-PDMS was spin-coated on the master mold at 1,000 rpm for 60 s, then baked at 70 ℃ for 2 h. A mixture of a 10:1 weight ratio of PDMS (Sylgard 184 A, Dow corning) and its curing agent (Sylgard 184 B, Dow corning) was poured on the *h*-PDMS layer and cured at 80 ℃ for 2 h. The cured soft mold was detached from the master mold, then used to replicate the nano-PER structure.

*Pre-treatment of the soft mold and the substrate*: Fluorosurfactant ((tridecafluoro-1,1,2,2-tetrahydrooctyl)trichlorosilane) is coated on the *h*-PDMS soft mold by vaporized coating at 130 ℃ for 5 min to decrease the surface tension of the soft mold. The glass substrate is also treated to increase the surface tension after the cleaning process using acetone, IPA, and DI water. At first, the cleaned glass is processed by $O_2$ plasma (CUTE-1MPR, Femto Science Inc.) at 100 W power and 100 sccm $O_2$ gas flow rate for 5 min. Then, the PMMA layer is spin-coated at 5,000 rpm for 1 min on the pre-treated glass substrate and heated at 180 ℃ for 5 min subsequently to fasten the uniform adhesive layer.

*Optical measurement*: A 532 nm laser (532 nm diode-pumped solid-state lasers, Thorlabs) was incident on a linear polarizer (Ø 1/2" unmounted linear polarizers, Thorlabs), half-wave plate (Ø 1/2" mounted achromatic half-wave plates, Thorlabs), and quarter-wave plate (Ø 1/2" mounted achromatic quarter-wave plates, Thorlabs) to form right circular polarized (RCP) light. A 500 μm diameter pinhole (P500HD - Ø 1/2" (12.7 mm) Mounted Pinhole, Thorlabs) was used to block the unnecessary light. A photodiode power sensor (S120C, Thorlabs) and a compact power and energy meter console (PM100D, Thorlabs) was used to measure the intensity of the light.

## Acknowledgements

This work was financially supported by the POSCO-POSTECH-RIST Convergence Research Center program funded by POSCO, the LGD-SNU incubation program funded by LG Display, and the National Research Foundation (NRF) grants (NRF-2019R1A2C3003129, CAMM-2019M3A6B3030637, NRF2019R1A5A8080290) funded by the Ministry of Science and ICT of the Korean government. J.K. and H.K. acknowledge the Alchemist fellowships from POSTECH. D.K. and Y.Y acknowledge the fellowships from the Hyundai Motor *Chung Mong-Koo* Foundation. Y.Y. acknowledges the NRF fellowship (NRF-2021R1A6A3A13038935) funded by the Ministry of Education of the Korean government.

**Appendix A. Supplementary data**

Supplementary data to this article can be found online.

*Supplementary Material*

# Metasurface Holography over 90% Efficiency in the Visible via Nanoparticle-Embedded-Resin Printing

Joohoon Kim,[1,†] Dong Kyo Oh,[1,†] Hongyoon Kim,[1] Gwanho Yoon,[2] Chunghwan Jung,[3] Jae Kyung Kim,[1] Trevon Badloe,[1] Seokwoo Kim,[1] Younghwan Yang,[1] Jihae Lee,[3] Byoungsu Ko,[1] Jong G. Ok,[4] and Junsuk Rho[1,3,5,6,*]

[1]Department of Mechanical Engineering, Pohang University of Science and Technology (POSTECH), Pohang 37673, Republic of Korea

[2]Department of Manufacturing Systems and Design Engineering, Seoul National University of Science and Technology, Seoul 01811, Republic of Korea

[3]Department of Chemical Engineering, Pohang University of Science and Technology (POSTECH), Pohang 37673, Republic of Korea

[4]Department of Mechanical and Automotive Engineering, Seoul National University of Science and Technology, Seoul 01811, Republic of Korea

[5]POSCO-POSTECH-RIST Convergence Research Center for Flat Optics and Metaphotonics, Pohang 37673, Republic of Korea

[6]National Institute of Nanomaterials Technology (NINT), Pohang 37673, Republic of Korea

[†]These authors contributed equally to this work.

*Corresponding author. E-mail: jsrho@postech.ac.kr

**Supplementary Data 1. The measured refractive index and extinction coefficient of the $TiO_2$ nano-PER film.**

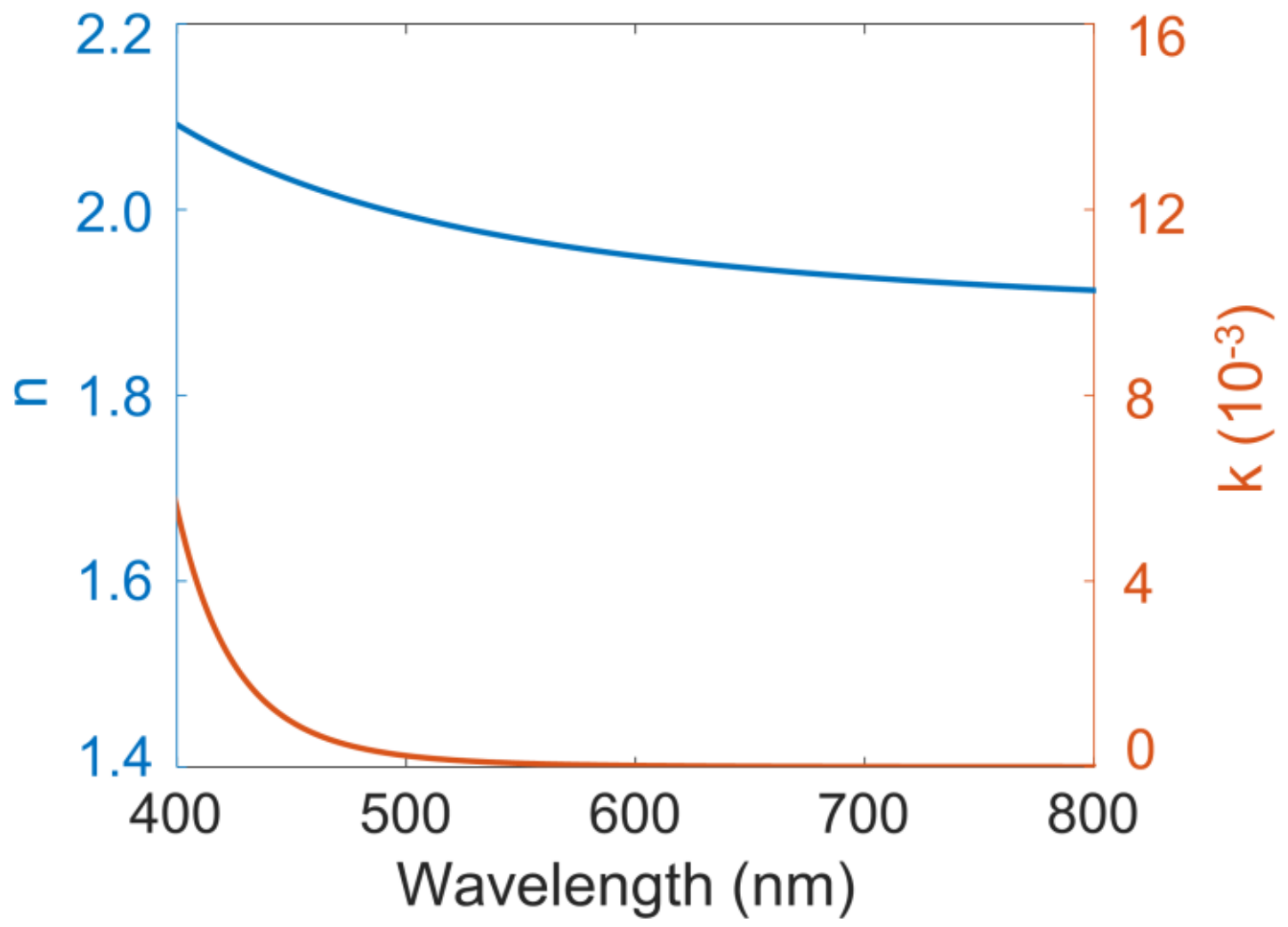

**Figure S1.** The measured refractive index and extinction coefficient of the nano-PER film with 80% $TiO_2$ weight ratio.

**Supplementary Data 2. Refractive index and simulated conversion efficiency with different $TiO_2$ weight ratio.**

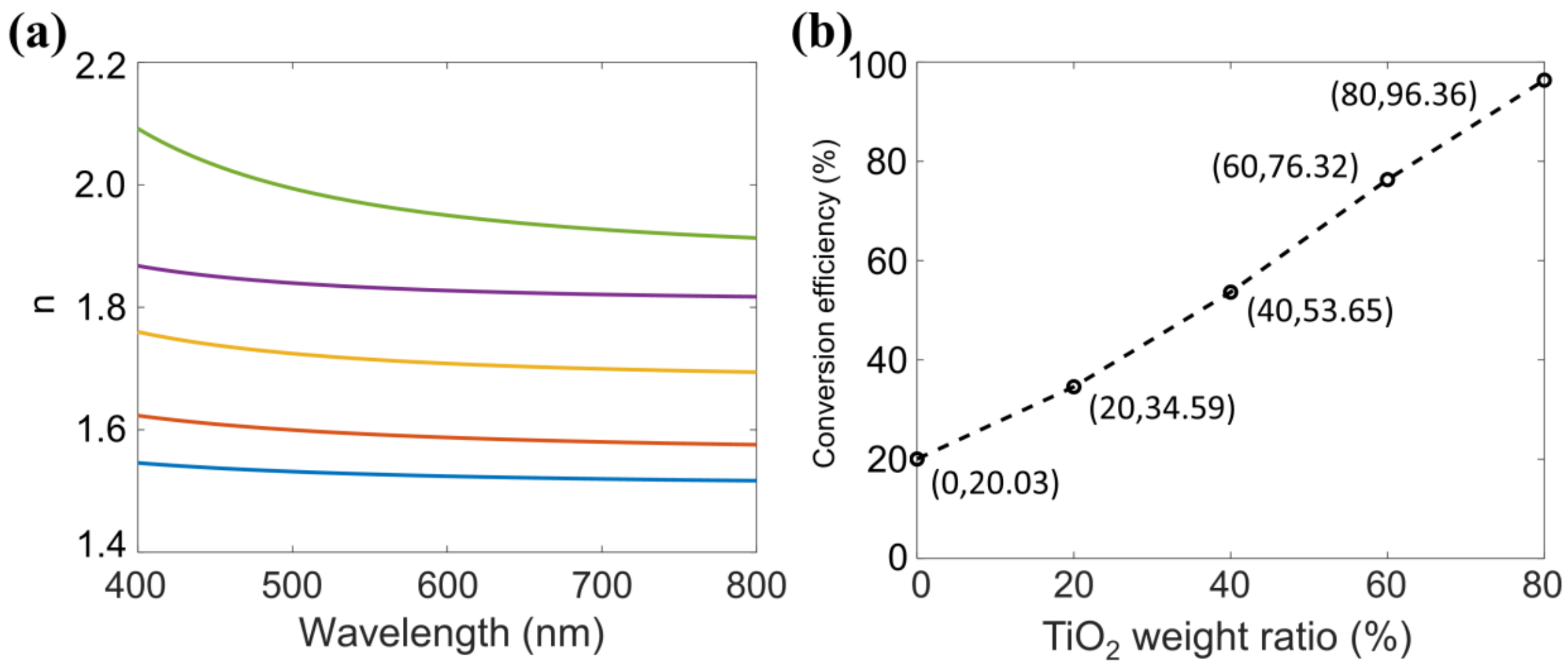

**Figure S2.** The measured refractive index and calculated conversion efficiency with different $TiO_2$ weight ratio. (a) Comparison of the refractive index of films with different $TiO_2$ weight ratio (blue: 0%; orange: 20%; yellow: 40%; purple: 60%; green: 80%) (b) Calculated conversion efficiency of the designed meta-atom at wavelength of 532 nm (period: 450 nm; height: 910 nm; length: 411 nm; width: 121 nm)

**Supplementary Data 3. The calculated refractive index using the Bruggeman mixing formula.**

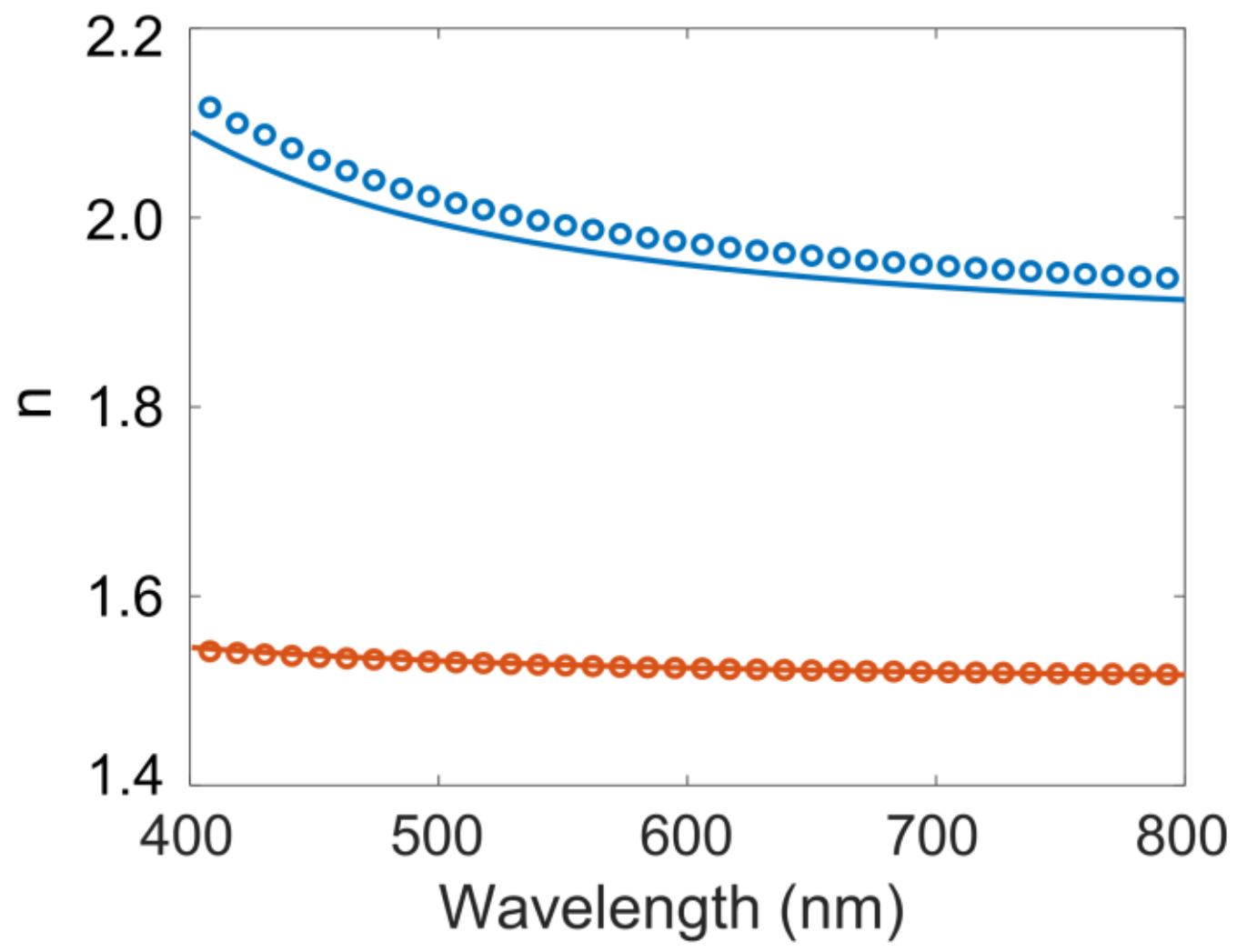

**Figure S3.** The refractive index of the typical resin (orange) and the nano-PER film with 80% ratio (blue); line: measured data using ellipsometry; circle: calculated data using the Bruggeman mixing formula

**Supplementary Data 4. Simulated conversion efficiency for different height meta-atoms.**

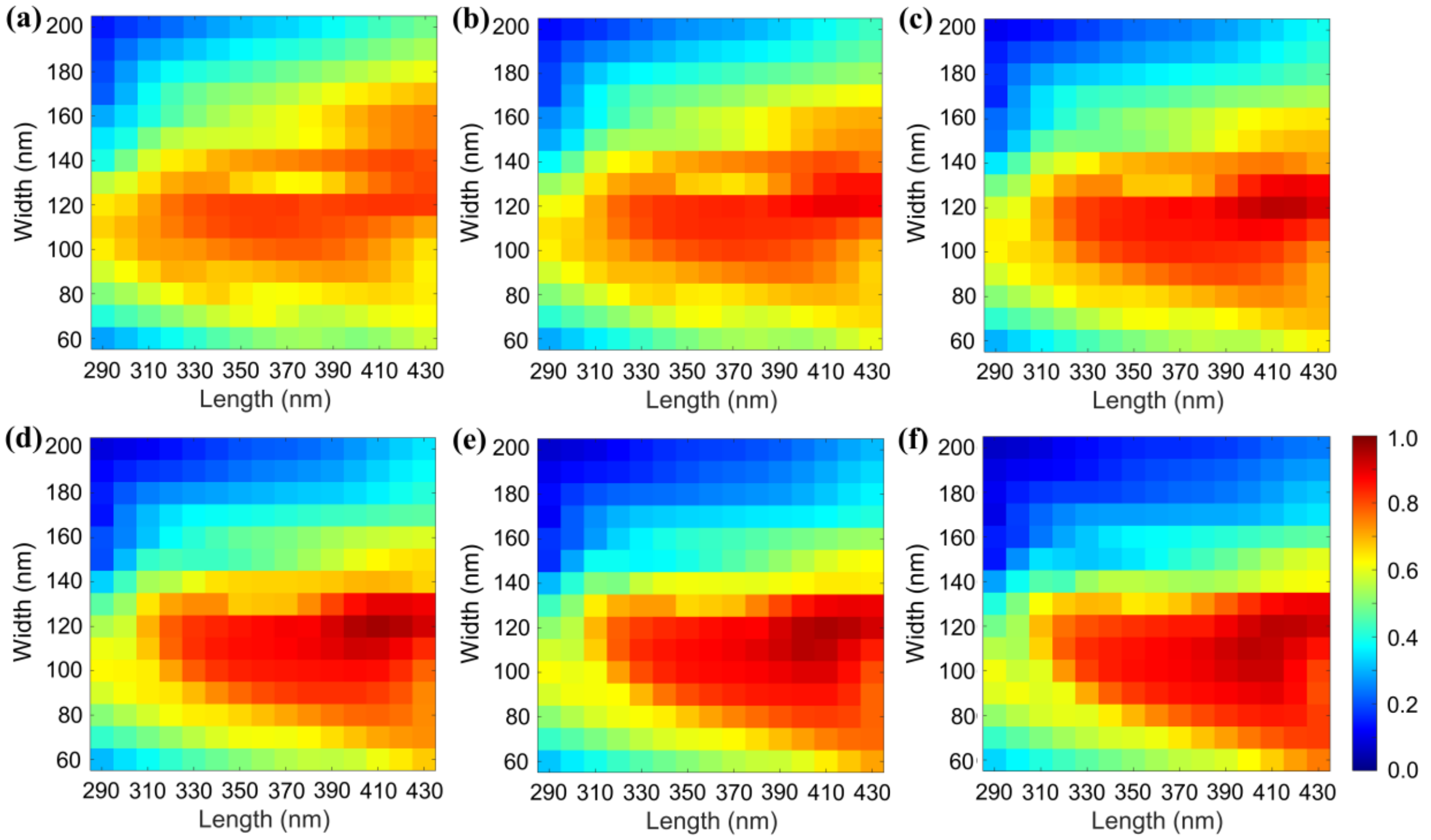

**Figure S4.** The calculated conversion efficiency for meta-atoms at 532 nm wavelength with a fixed period = 450 nm and different heights. The height of the meta-atoms is (a) 850 nm, (b) 870 nm, (c) 890 nm, (d) 910 nm, (e) 930 nm, and (f) 950 nm.

**Supplementary Data 5. Reflection analysis of metasurface**

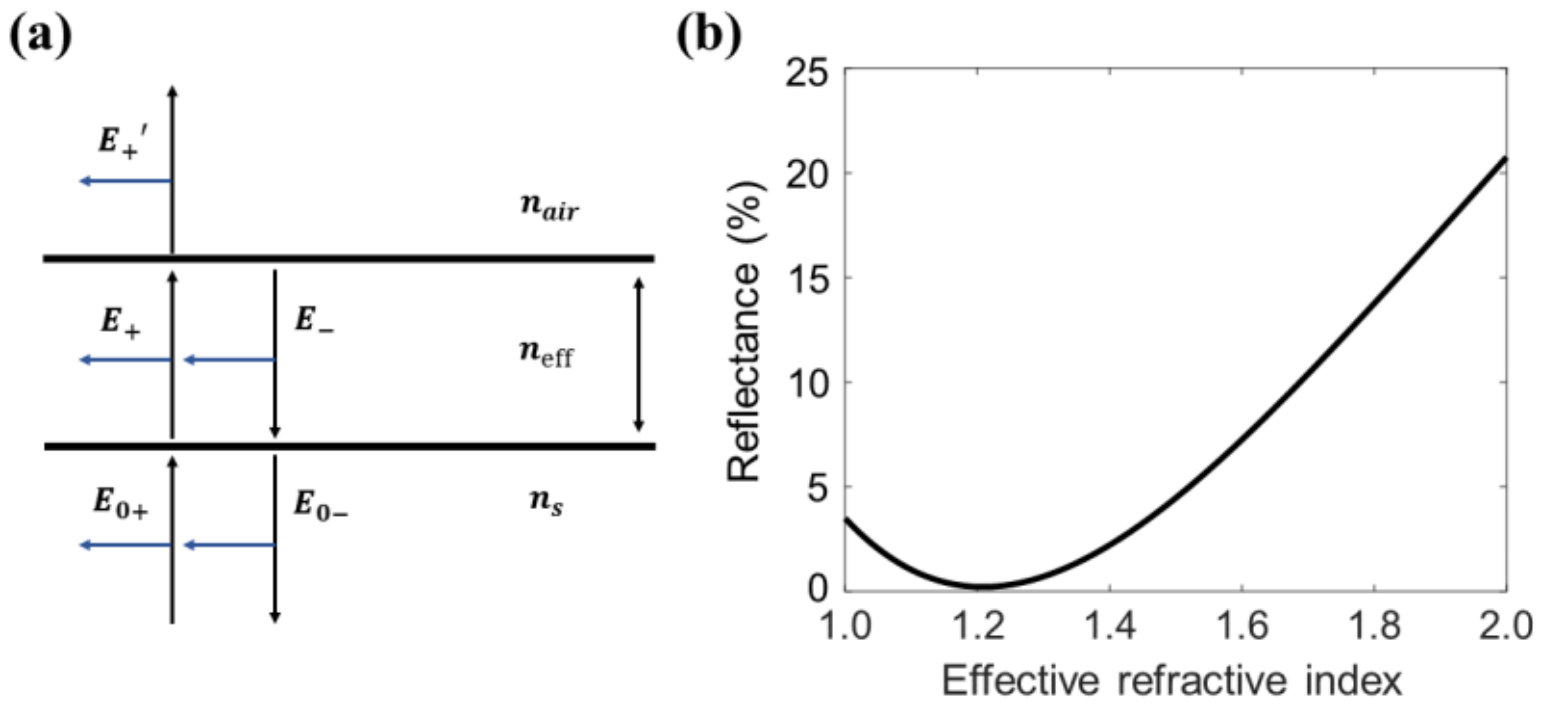

**Figure S5.** Reflection analysis of metasurface. (a) Schematic of electric field vectors for reflection at a film. (b) Calculated reflectance for wavelength of 532 nm and film height of 910 nm.

When light passes through the interface of two mediums with different refractive index, reflections inevitably occur, which can be explained by the Fresnel equations. According to the Fresnel equations, 3.5% of the light at 532nm wavelength is reflected when it passes from glass to air. Effective medium theory and Fresnel equations allow the analysis of reflections on metasurface by replacing metasurface with a single thin film that has a specific effective refractive index. At 532 nm wavelength, the refractive index of the material is about 1.98 and the refractive index the surrounding medium of air is 1, therefore the effective refractive index of the metasurface lies somewhere between the two values. We calculate the reflectivity according to the effective refractive index of the 910 nm height metasurface through the following equations (**Figure S5a**).

By the boundary condition, the transverse component of E, H field at the interface must be continuous.

At z = 0,

$$E_{0+} + E_{0-} = E_+ + E_- \tag{S1}$$

$$n_s(E_{0+} + E_{0-}) = n_{eff}(E_+ + E_-) \tag{S2}$$

At z = $w$,

$$E_+e^{ikw} + E_-e^{-ikw} = E_+'e^{ik_0w} \tag{S3}$$

$$n_{eff}(E_+e^{ikw} - E_-e^{-ikw}) = E_+'e^{ik_0w} \tag{S4}$$

where $k = n_{eff}k_0$ and $k_0$ is wave vector in free space.

(S1) and (S2) lead to

$$\begin{pmatrix} E_{0+} \\ E_{0-} \end{pmatrix} = \frac{1}{2}\begin{pmatrix} 1+n_r & 1-n_r \\ 1-n_r & 1+n_r \end{pmatrix} E_t \tag{S5}$$

where $\frac{n_{eff}}{n_s} \equiv n_r$

(S3) and (S4) lead to

$$\begin{pmatrix} E_{+} \\ E_{-} \end{pmatrix} = \frac{1}{2}\begin{pmatrix} \left(1+\frac{1}{n_{eff}}\right)e^{-ikw} \\ \left(1-\frac{1}{n_{eff}}\right)e^{ikw} \end{pmatrix} E'_{+}e^{ik_0 w} \tag{S6}$$

Finally, with the result (S5) and (S6), we have

$$\begin{pmatrix} E_{0+} \\ E_{0-} \end{pmatrix} = \frac{1}{4}\begin{pmatrix} 1+n_r & 1-n_r \\ 1-n_r & 1+n_r \end{pmatrix}\begin{pmatrix} \left(1+\frac{1}{n_{eff}}\right)e^{-ikw} \\ \left(1-\frac{1}{n_{eff}}\right)e^{ikw} \end{pmatrix} E'_{+}e^{ik_0 w} \tag{S7}$$

And the reflectance R is given by

$$\mathrm{R} = \left|\frac{E_{0-}}{E_{0+}}\right|^2 = \left|\frac{(1-n_r)\left(1+\frac{1}{n_{eff}}\right)e^{-ikw} + (1+n_r)\left(1-\frac{1}{n_{eff}}\right)e^{ikw}}{(1+n_r)\left(1+\frac{1}{n_{eff}}\right)e^{-ikw} + (1-n_r)\left(1-\frac{1}{n_{eff}}\right)e^{ikw}}\right|^2 \tag{S8}$$

The reflectance has a minimum point at the effective refractive index of 1.208, and the reflectance increases as the effective refractive index increase (**Figure S5b**). This minimum point is also consistent with the index that Lord Rayleigh proposed to have minimal reflection, which can be described as follows.[1]

$$n_1 = \sqrt{n_0 n_s} \tag{S9}$$

We also calculate the effective refractive index of our high efficiency metasurface using the Bruggmann mixing formula which can be expressed as follows.

$$\varepsilon_{\mathrm{BG}} = \frac{b \pm \sqrt{8\varepsilon_1\varepsilon_2 + b^2}}{4} \tag{S10}$$

where $b = (2f_1 - f_2)\varepsilon_1 + (2f_2 - f_1)\varepsilon_2$, $f_1$ is the filling factor of inclusion, $f_2$ is the filling factor of the host, $\varepsilon_1$ is the permittivity of inclusion, and $\varepsilon_2$ is the permittivity of the host. We confirm that the calculated effective refractive index of the metasurface, which is 1.209, matches well with the calculated anti-reflective index of refraction, which is 1.208. We confirmed that the metasurface is designed to reduce the reflection to increase efficiency.

**Supplementary Data 6. calculation of dispersion and polar surface tensions and work of adhesions between different interfaces.**

**Table S1.** Contact angles of various interfaces, and calculation of dispersion and polar surface tensions.

| Interfaces | Contact angle [°] | | $\gamma^d$ $[mJ/m^2]$ | $\gamma^p$ $[mJ/m^2]$ | $\gamma^{tot}$ $[mJ/m^2]$ |
|---|---|---|---|---|---|
| | Deionized Water | Diiodomethane | | | |
| Glass | 62.1 | 54.6 | 23.3 | 19.4 | 42.7 |
| Glass+ PMMA | 70.3 | 42.2 | 32.2 | 9.8 | 42.0 |
| Glass + $O_2$ plasma + PMMA | 71.3 | 38.3 | 34.6 | 8.4 | 43.0 |
| Glass + $O_2$ plasma | 4.1 | 43.4 | 22.8 | 49.8 | 72.6 |
| PER | 85.8 | 72.3 | 17.5 | 7.4 | 24.9 |
| PDMS mold | 94.5 | 64.1 | 24.4 | 2.0 | 26.4 |
| Fluorine-coated PDMS mold | 90.4 | 92.7 | 7.5 | 10.9 | 18.4 |

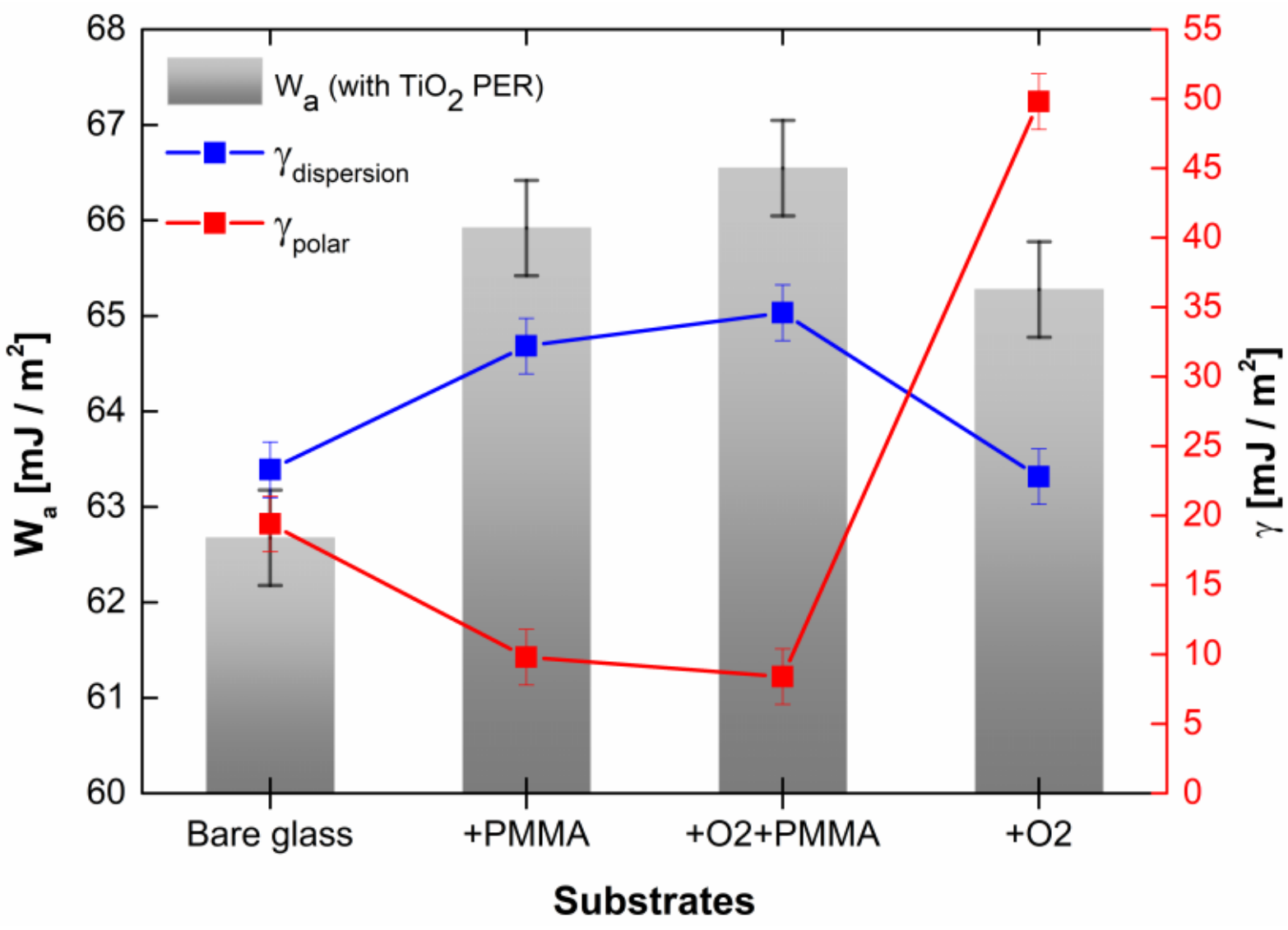

**Figure S6.** Calculation of work of adhesion ($W_a$) of various substrates with $TiO_2$ PER based on dispersion and polar components of their respective surface tensions.

**Supplementary Data 7. Captured image of the gradient palette hologram with low efficiency.**

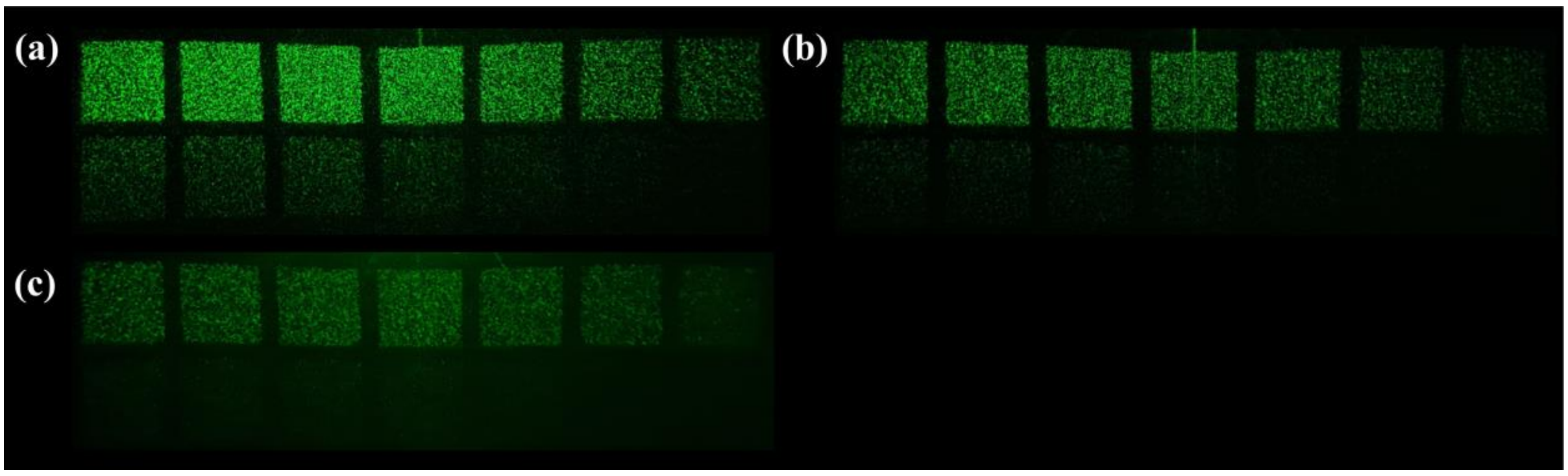

**Figure S7.** Captured image of the gradient palette hologram with conversion efficiencies of (a) 70%, (b) 50%, and (c) 30% .

**Supplementary Data 8. Setup to display hologram in the far-field.**

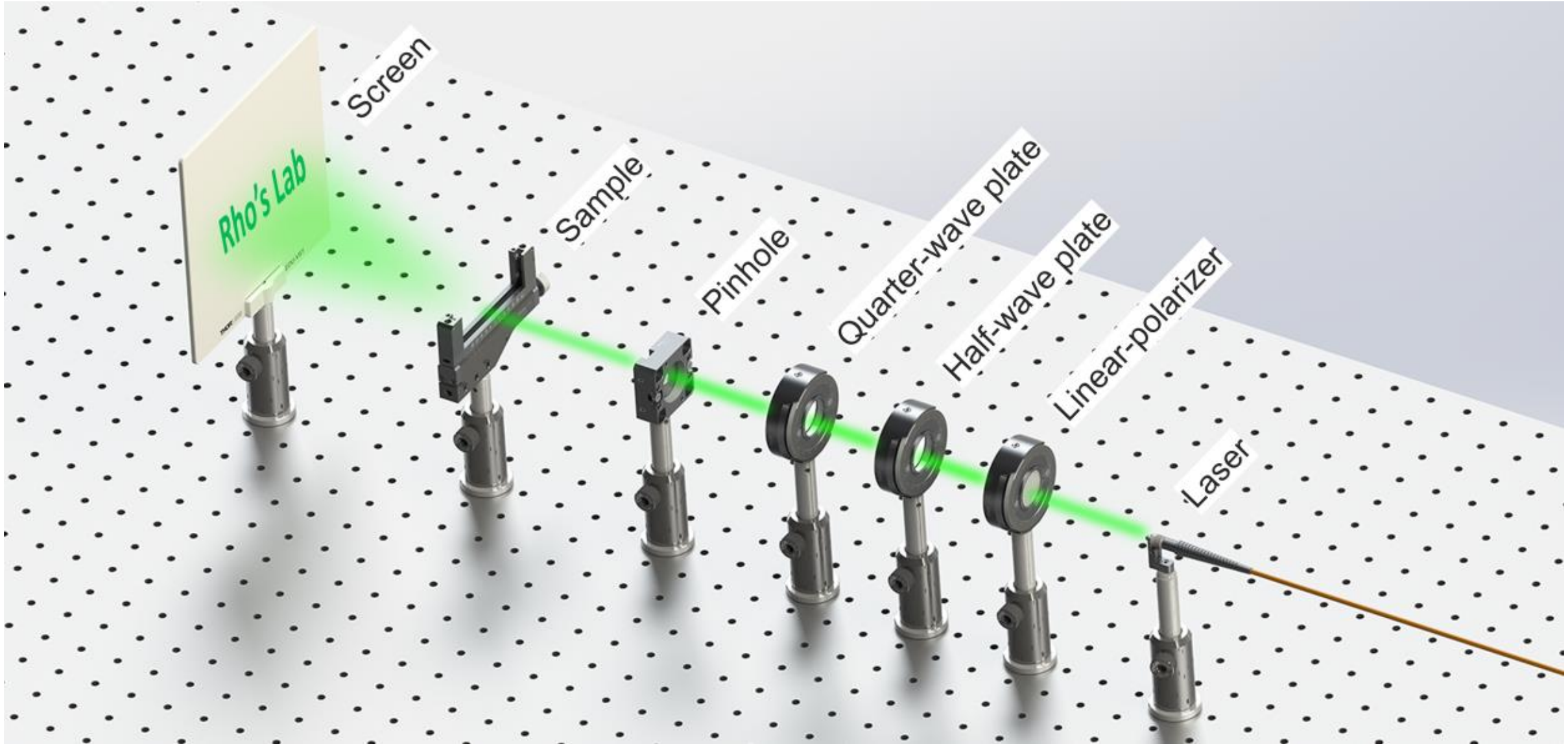

**Figure S8**. Schematic of the optical setup to display hologram in the far-field.

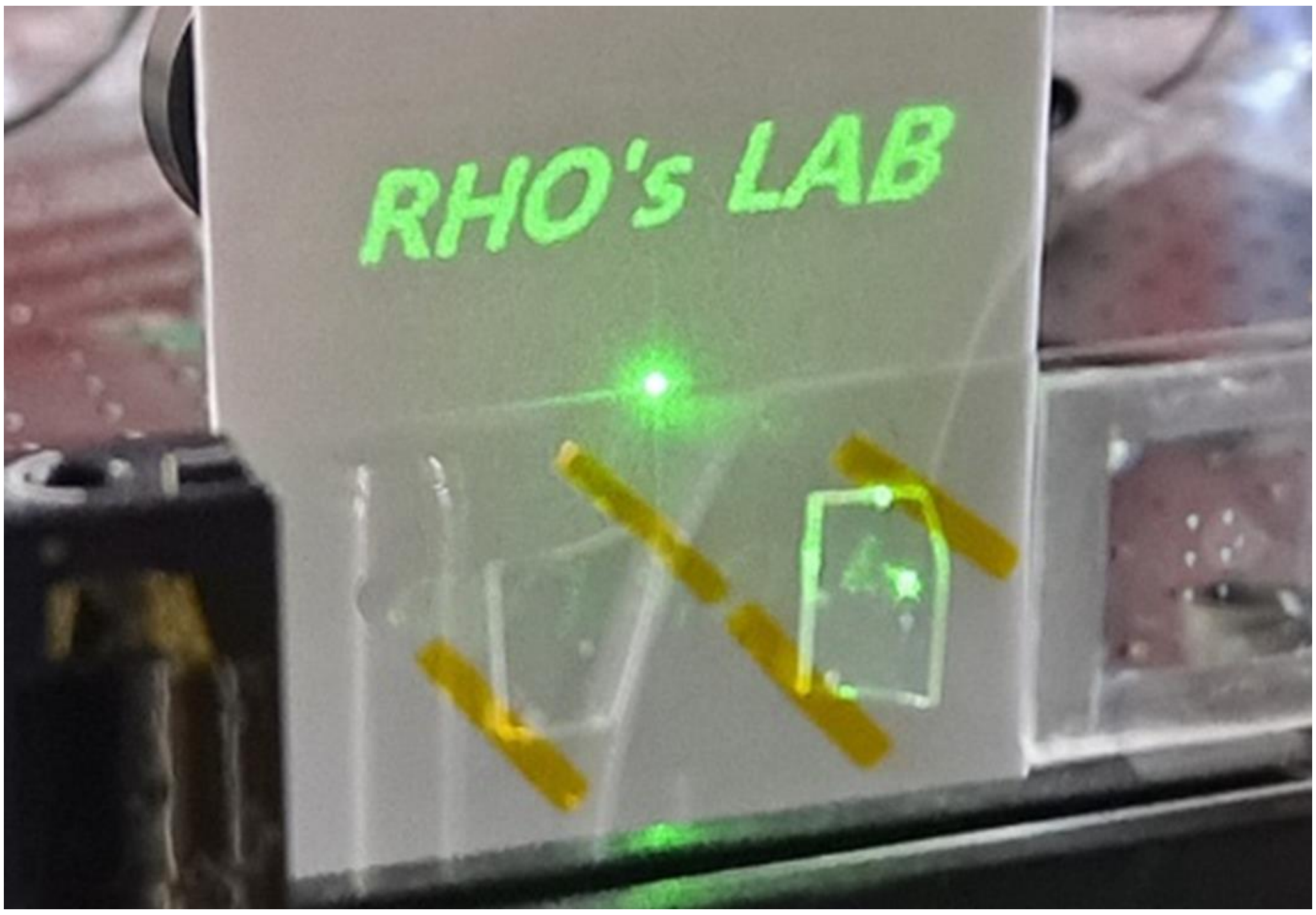

**Figure S9**. Captured image of the hologram under ambient light.

**Supplementary Data 9. Calculation of the measured conversion and zero-order efficiency.**

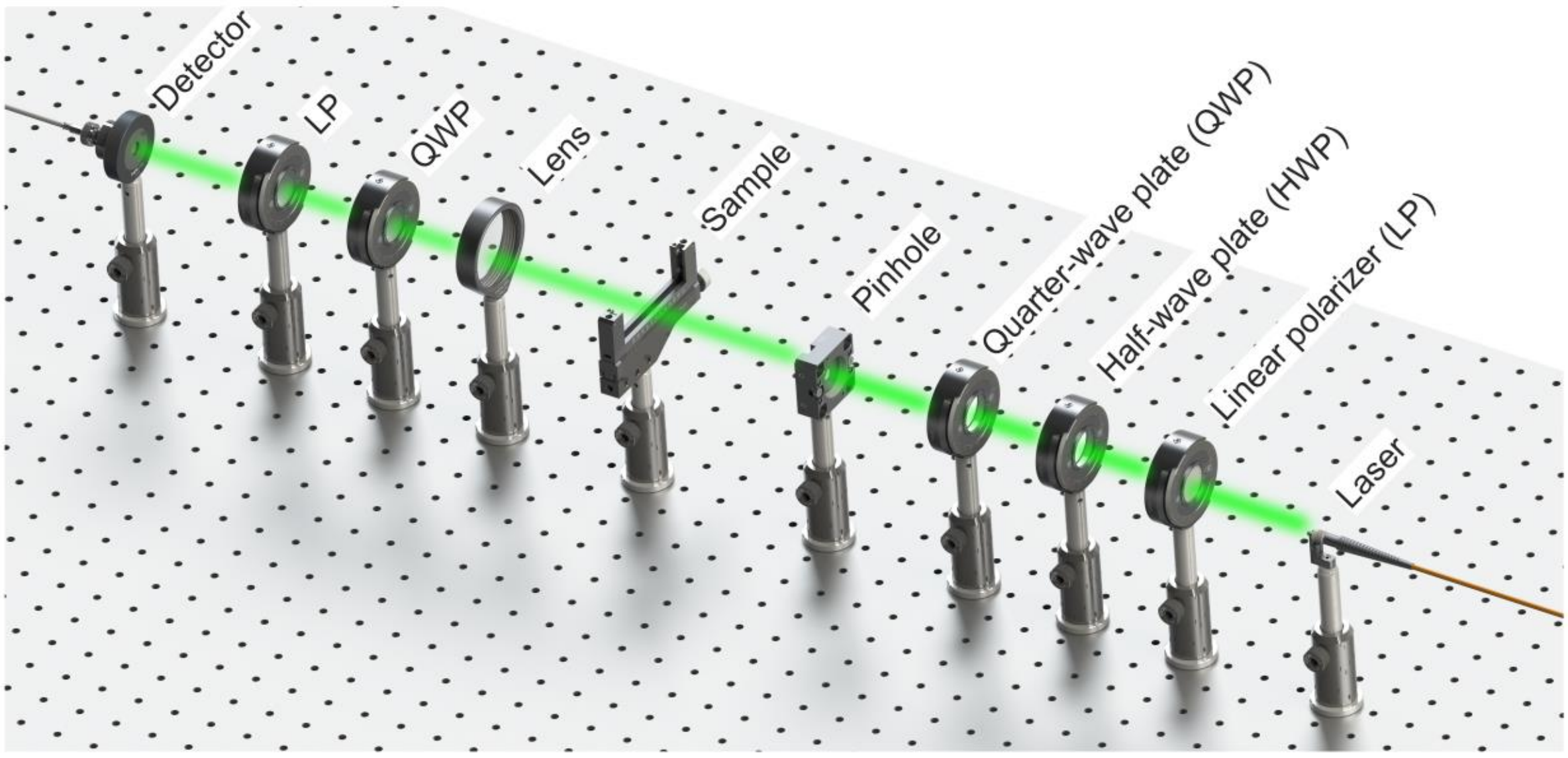

**Figure S10.** Schematic of the optical setup to measure the hologram efficiency.

**Table S2.** Data to calculate the conversion efficiency and zero-order efficiency of high efficiency nano-PER hologram.

| **Green (532 nm)** | Measured intensity with 500 µm pinhole (µW) | Calculated intensity for sample (µW) | Calculated efficiency (%) |
|---|---|---|---|
| Reference beam | 36.9 | 20.0 | - |
| Converted beam | 18.1 | 18.1 | 90.6 |
| Zero-order beam | 18.4 | 1.5 | 7.3 |

**Table S3.** Data to calculate the conversion efficiency and zero-order efficiency of the broadband nano-PER hologram.

| **Wavelength (450 nm)** | Measured intensity with 500 µm pinhole (µW) | Calculated intensity for sample (µW) | Calculated efficiency (%) |
|---|---|---|---|
| Reference beam | 12.77 | 6.9 | - |
| Converted beam | 3.5 | 3.5 | 51.0 |
| Zero-order beam | 7.1 | 1.3 | 18.6 |

| **Wavelength (475 nm)** | Measured intensity with 500 µm pinhole (µW) | Calculated intensity for sample (µW) | Calculated efficiency (%) |
|---|---|---|---|
| Reference beam | 18.8 | 10.2 | - |
| Converted beam | 5.5 | 5.5 | 54.3 |
| Zero-order beam | 10.7 | 2.1 | 20.5 |
| **Wavelength (500 nm)** | Measured intensity with 500 µm pinhole (µW) | Calculated intensity for sample (µW) | Calculated efficiency (%) |
| Reference beam | 15.2 | 8.3 | - |
| Converted beam | 4.8 | 4.8 | 57.9 |
| Zero-order beam | 7.7 | 0.7 | 8.9 |
| **Wavelength (525 nm)** | Measured intensity with 500 µm pinhole (µW) | Calculated intensity for sample (µW) | Calculated efficiency (%) |
| Reference beam | 30.2 | 16.4 | - |
| Converted beam | 13.1 | 13.1 | 80.0 |
| Zero-order beam | 15.4 | 1.6 | 9.6 |
| **Wavelength (550 nm)** | Measured intensity with 500 µm pinhole (µW) | Calculated intensity for sample (µW) | Calculated efficiency (%) |
| Reference beam | 42.4 | 23.0 | - |
| Converted beam | 18.6 | 18.6 | 81.0 |
| Zero-order beam | 21.6 | 2.2 | 9.4 |
| **Wavelength (575 nm)** | Measured intensity with 500 µm pinhole (µW) | Calculated intensity for sample (µW) | Calculated efficiency (%) |
| Reference beam | 24.4 | 13.2 | - |
| Converted beam | 8.1 | 8.1 | 61.4 |
| Zero-order beam | 13.7 | 2.5 | 19.2 |
| **Wavelength (600 nm)** | Measured intensity with 500 µm pinhole (µW) | Calculated intensity for sample (µW) | Calculated efficiency (%) |
| Reference beam | 19.9 | 10.8 | - |
| Converted beam | 6.2 | 6.2 | 57.4 |
| Zero-order beam | 12.0 | 2.9 | 26.9 |
| **Wavelength (625 nm)** | Measured intensity with 500 μm pinhole (μW) | Calculated intensity for sample (μW) | Calculated efficiency (%) |
| Reference beam | 21.8 | 11.8 | - |
| Converted beam | 6.9 | 6.9 | 58.5 |
| Zero-order beam | 13.8 | 3.8 | 32.1 |
| **Wavelength (650 nm)** | Measured intensity with 500 μm pinhole (μW) | Calculated intensity for sample (μW) | Calculated efficiency (%) |
| Reference beam | 124.5 | 67.4 | - |
| Converted beam | 40.2 | 40.2 | 59.6 |
| Zero-order beam | 82.5 | 25.4 | 37.7 |

We measure the intensity of the light using a 500 μm pinhole to calculate the conversion and zero-order efficiency. The fabricated sample is a square with a side length of 296 μm. We use a pinhole with a diameter of 500 μm, so that the sample can be fully covered. However, we should only consider the light that passed through the sample. Therefore, we calculate the intensity of light that does not pass through the sample by dividing the area of the sample with the area of the pinhole. This is then subtracted from the reference beam and the zero-order beam. Finally, we calculate the conversion and zero-order efficiency by dividing the intensity of the converted and zero-order beams by the reference beam, respectively.

**Supplementary Data 10. Comparison reported high efficiency metaholograms and metalens using PB phase.**

**Table S4.** Comparison reported high efficiency metaholograms and metalenses using PB phase. Efficiency is defined as the ratio of the imaging optical power to the total incident optical power.

| Reference | Type (Application) | Material | Wavelength (nm) | Theoretical efficiency (%) | Experimental efficiency (%) |
|---|---|---|---|---|---|
| This work | Hologram | $TiO_2$ nano-PER | 532 | 96.4 | 90.6 |
| [2] | Hologram | a-Si:H | 633 | 74 | 61 |
| [3] | Lens | $TiO_2$ | 405 | 95 | 86 |
| [4] | Lens | c-Si | 532 | 71 | 67 |
| [5] | Lens | GaN | 532 | - | 91.6 |
| [6] | Hologram | $TiO_2$ | 480 | - | 82 |

**Supplementary Data 11. Calculation of the cost of single sample.**

**Table S5**. Breakdown of the cost of a single sample.

| **Material** | Dipentaerythritol penta-/hexa-acrylate | 1-Hydroxycyclohexyl phenyl ketone | MIBK | DT-TiOA-30MIBK (N30) |
|---|---|---|---|---|
| **Role** | Monomer | Photo-initiator | Solvent | $TiO_2$ NP- dispersed MIBK |
| **Supplier** | Sigma-Aldrich | Sigma-Aldrich | Duksan general science | Ditto technology |
| **Volume (or weight)** | 100 mL | 50 g | 1 L | 200 g |
| **Price (USD)** | 80 | 70 | 9 | 175 |
| **Volume (or weight) per sample** | 0.1 mL | 0.85 g | 0.7 mL | 0.104 g |
| **Price per sample (USD)** | 0.08 | 1.19 | 0.0063 | 0.091 |

Once the soft mold is fabricated, high-efficiency holograms made up of the $TiO_2$ nano-PER can be printed repeatedly. We verify cost of a single sample using the proposed fabrication method by breaking down the cost of constituent parts of the $TiO_2$ nano-PER. Based on the purchase history of the materials used to produce the $TiO_2$ nano-PER, the total cost for single sample fabrication is calculated to be around 1.37 USD.

**Supplementary references:**